\documentclass[twocolumn,amsmath,amssymb]{revtex4} 
\usepackage{graphicx}
\usepackage{dcolumn}
\usepackage{bm}
\begin{document} 
 
\title{Taking stock of the quantum Hall effects: Thirty years on} 
\author{Tapash Chakraborty$^\ddag$ and Klaus von Klitzing$^\ast$} 
\affiliation{$^\ddag$Department of Physics and Astronomy, 
University of Manitoba, Winnipeg, Canada R3T 2N2} 
\affiliation{$^\ast$Max Planck Institut f\"ur Festk\"orperforschung, 
Heisenbergstrasse 1, D-70569 Stuttgart, Germany} 
 
\date{\today} 
\begin{abstract} 
The quantum Hall effects, discovered about thirty years ago have remained  
one of the most spectacular discoveries in condensed matter physics in the  
past century. Those discoveries triggered huge expansion in the field  
of low-dimensinal electronic systems, the area grew at an unprecedented 
rate and continues to expand. Novel and challenging observations, be it  
theoretical or experimental, have been reported since then on a regular  
basis.  Additionally, the effects have inspired physicists to find analogous  
situations in far-flung fields as disparate as string theory or black hole  
physics. 
 
\end{abstract} 
\maketitle 
 
The quantum Hall effects (QHEs) are now about 30 years old. The date of birth  
of the original effect was duely recorded as February 5, 1980 at around  
2 a.m. during an experiment at the High Magnetic Field Laboratory in  
Grenoble, France \cite{discovery}, while its fractional counterpart was  
discovered on October 7, 1981 at the Francis Bitter Magnet Laboratory,  
Massachusetts, USA \cite{site}. The objectives of the Grenoble  
experiment were to answer some of the fundamental questions in the  
electronic transport of silicon field effect transistors, such as, how  
can one improve the mobility of these devices, or what are the dominant  
scattering processes in the dynamics of electrons at the nanometer scale  
at the interface between silicon and silicon dioxide. Specially designed  
devices (Hall devices), such as the one shown in Fig.~1, which allow  
direct measurement of the resistivity tensor were considered. Low  
temperatures (typically 4.2 K) were used so that the scattering processes  
involving electron-phonon interactions were suppressed. Application of a  
magnetic field was an already established method to gather information  
about the microscopic details of a semiconductor \cite{ando_review}. 
 
\begin{figure} 
\begin{center}\includegraphics[width=7cm]{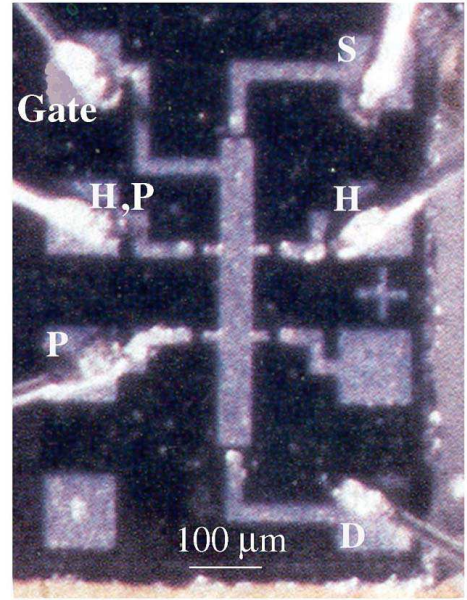} 
\end{center} 
\caption{A typical silicon MOSFET device used for measurements of the Hall  
effect. 
} 
\label{Fig_pub_1} 
\end{figure} 
 
It was known since 1966 that electrons accumulated at the surface of a silicon  
crystal by a positive voltage at the gate (i.e., a metal plate parallel to the  
surface) form a two-dimensional electron gas (2DEG) \cite{fowler}. The energy  
for electron motion perpendicular to the surface is quantized (dimensional 
quantization), while in the presence of a perpendicular magnetic field, the  
motion of electrons in the plane is also quantized (Landau quantization). In the  
ideal case, the energy spectrum of a 2DEG in strong magnetic fields consists of  
discrete Landau energy levels (normally broadened due to impurity scattering)  
with an equal energy spacing. The QHE is observed if the Fermi energy lies in  
the energy gap and if the temperature is so low that excitations across the gap  
are not possible.  
 
The experimental results that led to the discovery of the QHE are shown in  
Fig.~2. The blue curve is the electrical resistance of the silicon field  
effect transistor as a function of the gate voltage. Since the electron  
concentration increases linearly with increasing gate voltage, the electrical  
resistance decreases monotonically. Further, the Hall voltage (if a constant  
magnetic field of, say, 19.8 Tesla is applied) decreases with increasing gate  
voltage, since it is inversely proportional to the electron concentration.  
The black curve shows the Hall resistance with a clear plateau at a gate  
voltage where the longitudinal resistance vanishes. The uniqueness of these  
findings was that the Hall plateau can be expressed with high precision as  
$\rho_{xy}=h/ne^2$ ($h$ is the Planck constant, $e$ is the elementary charge,  
and $n$ is the number of fully occupied Landau levels).  
  
\begin{figure} 
\begin{center}\includegraphics[width=7cm]{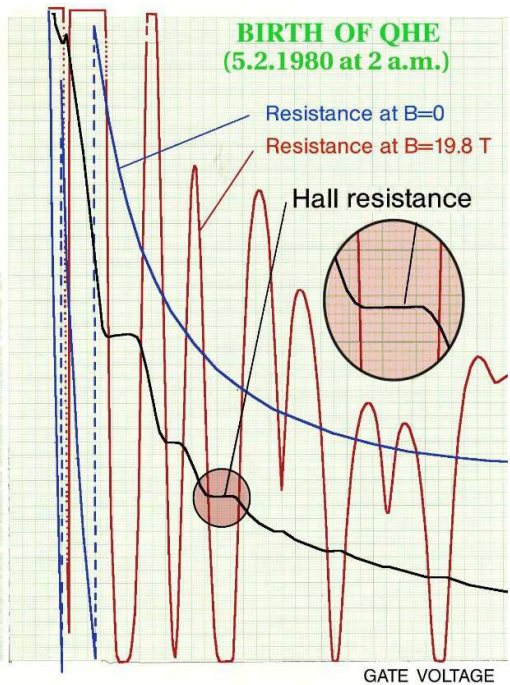} 
\end{center} 
\caption{Hall resistance and longitudinal resistance (at B=0 and B=19.8 Tesla)  
of a silicon MOSFET at liquid helium temperature versus the gate voltage. The  
enlarged part depicts the Hall plateau at filling factor 4.} 
\label{Fig_pub_2} 
\end{figure} 
 
The epoch making discovery was the `exact quantization' of Hall resistance  
to a fundamental value of $h/e^2=25812.807 ...$ Ohm that is incredibly robust.  
This value is independent of the material, geometry and microscopic details  
of the semiconductor \cite{klitzing_PRL,nobel}. Measurements of the Hall  
conductance have been found to be integral multiples of $e^2/h$ to nearly  
one part in a billion. This has facilitated the definition of a new practical  
standard for electrical resistance based on the resistance quantum given by  
the {\it von Klitzing constant} $R_K=25812.807449\pm0.000086$ Ohm. Since 1990,  
a fixed value of $R_{K-90}=25,812.807$ Ohm has been adopted internationally  
as a standard for resistance calibration \cite{quinn,jeanneret} (Table~1).  
Recent discussions about a new definition of the units of measurements based on 
fundamental constants led to the recommendation \cite{recent_metro} to fix not  
only the value of the velocity of light but also the values of the Boltzmann  
constant, the Avogadro constant, the Planck constant and the elementary charge,  
which automatically means that also the von Klitzing constant will be a fixed  
number within such a new system of international units (SI system). Within the  
present SI system the QHE provides an extremely precise independent determination  
of the fine structure constant which is ``one of the fundamental constants of nature  
characterizing a whole range of physics including elementary particle,  
atomic, mesoscopic and macroscopic systems" \cite{kinoshita}.  
 
In a simple minded picture to explain the observed step-like behaviour of  
resistivity, one could begin with the non-interacting electron system in a  
perpendicular magnetic field. The Landau levels (LLs) are known to be highly  
degenerate, with degeneracy defined as the number of states per unit area,  
$eB/h$. As each of the degenerate states is filled, fewer states remain unoccupied  
and the resistivity decreases. Once the LL is completely filled, there remains a 
gap to the next energy LL and the resistivity vanishes at sufficiently low 
temperatures. Due to the presence of impurities in the sample, there are  
localized states that can be filled but they do not contribute to the conductivity.  
The remarkable precision of Hall quantization which is oblivious to the material  
characteristics, impurities, and different geometries, was attributed to the  
subtle manifestation of the principle of gauge invariance \cite{laughlin_integer}. 
  
The QHEs are characterized by the filling factor $\nu$ ($\nu=$total number of 
electrons/number of flux quanta passing through the sample=$n_s\Phi_0/B$, where 
$n_s$ is the carrier density, $\Phi_0=h/e$ is the flux quantum and $B$ is the 
magnetic field). The integer QHE (IQHE) corresponds to $\nu$ being a simple 
integer. In 1982, Tsui, St\"ormer and Gossard discovered \cite{tsg} that in  
devices with much less disorder, the QHE appears with $\nu$ having rational 
fractional values (Fig.~3). This fractional QHE (FQHE) arises purely  
due to electron-electron interactions. The original observation of a FQHE at 
$\nu=\frac13,\frac23$ was superbly described by Laughlin \cite{laughlin,TC_book} 
who introduced a many-body wavefunction that was based on an inspired guess.  
It was confirmed subsequently by various numerical studies \cite{TC_book}. 
The novelty of the Laughlin state was that, it described an incompressible 
state of the electron liquid whose low-energy excitations are fractionally-charged 
quasiparticles and quasiholes \cite{laughlin,TC_book}, not unlike quarks 
\cite{greiter}. They also obey fractional statistics \cite{halperin,wilczek}, 
which means that the interchange of two such objects multiplies the wavefunction by 
a phase which may take any value (``anyons") instead of just +1 (bosons) 
or $-1$ (fermions). The QHE depends crucially on the existence of a gap in the 
excitation spectrum. In the case of the IQHE, the gaps are the single-particle  
type kinetic energy gap between Landau levels, the spin gap, and the valley gap in 
Si. The gap in the FQHE on the other hand, arises purely from electron-electron  
interactions. Electron spin was also found to play an important role in the FQHE  
ground state and excitations ({\it spin-reversed quasiparticles}) \cite{TC_book,TC_spin}. 
  
\vspace*{0.5cm} 
\noindent 
Table~1: Summary of high precision data for the quantized Hall resistance  
until 1988 which led to the fixed value of 25818.807 Ohm recommended as a  
reference standard for all resistance calibrations after January 1, 1990.  
\vspace*{0.2cm} 
 
\begin{tabular}{ll} 
{\bf (Hall-) Resistance} & ${\cal R}_{\rm H}$ \\ 
\hline\\ 
PRL 45, 494 (1980) & 25 812.68 (8) $\Omega$ \\ 
BIPM (France) & 25 812.809 (3) $\Omega$ \\ 
PTB (Germany) & 25 812.802 (3) $\Omega$ \\ 
ETL (Japan) & 25 812.804 (8) $\Omega$ \\ 
VSL (The Netherlands) & 25 812.802 (5) $\Omega$ \\ 
NRC (Canada) & 25 812.814 (6) $\Omega$ \\ 
EAM (Switzerland) & 25 812.809 (4) $\Omega$ \\ 
NBS (USA) & 25 812.810 (2) $\Omega$ \\ 
NPL (UK) & 25 812.811 (2) $\Omega$ \\ 
{\bf 1.1. 1990} & {\bf 25 812.80700} $\Omega$\\ \hline 
\end{tabular} 
\vspace*{0.5cm} 
 
When better quality samples started revealing more and more filling fractions 
\cite{stormer}, it soon became clear that Laughlin's theory was inadequate to 
describe those higher-order filling factors. In the composite fermion 
picture \cite{jain_book}, trial wave functions for ground states and excitations 
were introduced that correctly predict the most prominent observed FQH states.  
In contrast to the observation of the FQHE at odd-denominator filling factors, 
for most even-denominator fractions, and in particular at $\nu=\frac12$, no 
effect has been observed. According to the fermion-Chern-Simon picture \cite{HLR}, 
the system at $\nu=\frac12$ is compressible and is not expected to display 
any QHE. Exactly at $\nu=\frac12$ and within the mean-field approximation, the 
actual electron system becomes equivalent to a gas of fermions in zero magnetic  
field. There are experimental indications in support of this idea \cite{willet}. 
One surprising discovery in the FQHE was the strongly correlated electronic  
state at the half-filled second orbital $(\nu=\tfrac52$) Landau level \cite{jim}.  
This observation surely did not tally with the existing theories for all other 
FQH states. It has been proposed theoretically that this filling fraction  
corresponds to degenerate ground states and fractionally-charged non-Abelian quasiparticles  
\cite{read}. Interchange of two quasiparticles of this type would shift the system  
between orthogonal ground states. This state is proposed to have properties  
appropriate for fault tolerant quantum computation \cite{qubit}. The non-Abelian  
nature of the $\frac52$ state is yet to be directly confirmed by experiments.  
However, in recent experiments \cite{jacoby}, the quasiparticle charge was determined  
to be $e^*_{\frac52}= e/4$, in agreement with the proposed paired FQH state at  
$\nu=\frac52$ \cite{read}. This property is different from all other observed FQH states. 
  
\begin{figure} 
\begin{center}\includegraphics[width=9cm]{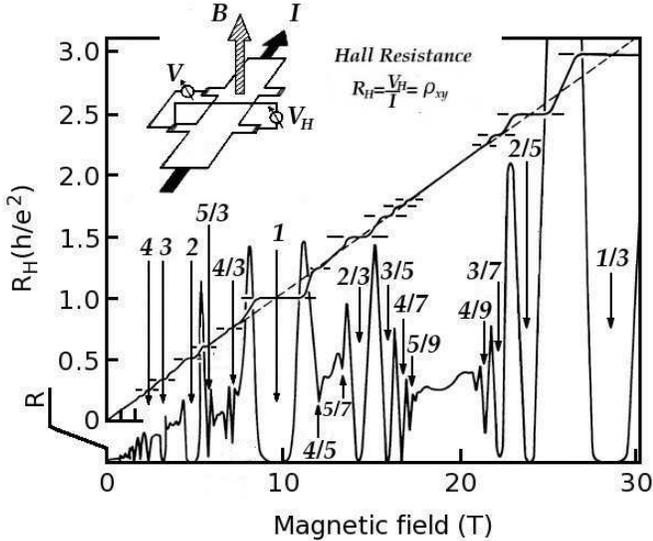} 
\end{center} 
\caption{Fractional (and integer) filling factors where QHE is observed 
(adopted from \cite{TC_book}). 
} 
\label{Fig_pub_3} 
\end{figure} 
 
The theory of Laughlin which introduced several novel concepts in correlated 
quantum fluids, inspired analogous effects in other subfields of physics. 
The QHE was generalized to four dimensions \cite{four_dimen} in order to  
study the ``interplay between quantum correlations and dimensionality in  
strongly correlated systems". Two-dimensional electron systems were modeled  
by strings interacting with D-branes \cite{susskind}. Here the fractionally-charged  
quasiparticles and composite fermions were described in the language of 
string theory. An interesting analogy between the QHE and black hole has been 
reported, and in particular, the edge properties of a QHE system have been used to model 
black hole physics from the point of view of an external observer \cite{black_hole}.  
Important developments of the QHE have also taken place from 
the field theoretical point of view \cite{ezawa}.  
 
There has been a lot of excitement recently about a new state of matter, the  
topological insulator \cite{kane}, which has a bulk insulating gap, but 
gapless electronic states (topologically protected against scattering by 
time-reversal symmetry) on the sample boundary. In two-dimensions the  
topological insulator is a quantum spin Hall system, somewhat akin to the 
IQH state. Finally, the QHE has played a crucial role in a novel two-dimensional  
system discovered recently, graphene. The latter is a single-atom-thick layer of  
carbon atoms arranged in a hexagonal lattice with remarkable attributes 
\cite{geim,my_pic}. Charge carriers in graphene behave as massless Dirac  
fermions, whose dynamics is governed by the Dirac equation. The quantization 
condition of Hall resistance in graphene is different from that in a conventional  
2DEG by a half-integer shift \cite{graphene_QHE}, and has been reliably measured 
even at room temperature \cite{novo_QHE} which is attributed to large 
cyclotron gaps of Dirac fermions in graphene. The fractional QHE in 
graphene was studied theoretically \cite{vadim_fqhe} and was subsequently  
observed \cite{fqhe_kim}. 
 
The QHEs are truly remarkable macroscopic quantum phenomena observed in  
two-dimensional electron systems. Discovery of IQHE was clearly just the  
beginning of a long sequence of discoveries in this field. Experiments on  
the QHE continue to reveal a countless number of often unexpected and  
challenging results. Theorists have been busy developing novel concepts in  
order to deal with these phenomena. What is happening now in the field of  
low-dimensional electron systems is nothing short of a revolution  
that shows no signs of running out of steam in the immediate future. 
 
One of us (T.C.) would like to thank Peter Maksym (Leicester, UK) for a 
critical reading of the manuscript.

\end{document}